\RequirePackage{fix-cm}
\documentclass{svjour3}                     

\smartqed  
\usepackage{graphicx}

\usepackage{wrapfig}
\usepackage{soul}

\journalname{Climatic Change}

\begin{document}

\title{Deepening of the ocean mixed layer at the northern Patagonian continental shelf: a numerical study}

\titlerunning{Deepening of the ocean mixed layer at the northern Patagonian continental shelf}

\author{Juan Zanella  \and
	Ezequiel \'Alvarez \and \\
	Andr\'es Pescio \and
	Walter Dragani}

\institute{J. Zanella \at
              Departamento de F\'\i sica, FCEyN, Universidad de Buenos Aires, Ciudad Universitaria, Pabell\'on I, (1428) Ciudad Aut\'onoma de Buenos Aires, Argentina \\
              Tel.: (54-11) 4576-3353\\
              Fax: (54-11) 4576-3357\\
              \email{zanellaj@df.uba.ar}
           \and
           E. \'Alvarez \at
              Departamento de F\'\i sica, FCEyN, Universidad de Buenos Aires, Ciudad Universitaria, Pabell\'on I, (1428) Ciudad Aut\'onoma de Buenos Aires, Argentina
           \and
           A. Pescio \at
           Servicio de Hidrograf\'\i a Naval, Ministerio de Defensa, Av. Montes de Oca 2124, (C1270ABV) Ciudad Aut\'onoma de  Buenos Aires, Argentina
            \and
           W. Dragani \at
           UMI IFAECI/CNRS - CONICET - UBA, Ciudad Universitaria, Pabell\'on II, 2do. Piso, (C1428EGA) Ciudad Aut\'onoma de Buenos Aires, Argentina; and Servicio de Hidrograf\'\i a Naval, Ministerio de Defensa, Av. Montes de Oca 2124, (C1270ABV) Ciudad Aut\'onoma de  Buenos Aires, Argentina
}

\date{Received: date / Accepted: date}

\maketitle

\begin{abstract}
A possible deepening of the ocean mixed layer was investigated at a selected point of the Patagonian continental shelf where a significant positive wind speed trend was estimated. Using a 1-dimensional vertical numerical model (S2P3) forced by atmospheric data from NCEP/NCAR I reanalysis and tidal constituents from TPXO 7.2 global model on a long term simulation (1979-2011), it was found that the mixed layer thickness presents a significant and positive trend of $10.1$ $\pm$ $1.4$ cm/yr. Several numerical experiments were carried out in order to evaluate the impact of the different atmospheric variables (surface zonal and latitudinal wind components, air temperature, atmospheric pressure, specific humidity and cloud coverage) considered in this study. As a result it was found that an increase in the wind speed can be considered as the main responsible of the ocean mixed layer deepening at the selected location of the Patagonian continental shelf. A possible increasing in the mixed layer thickness could be directly impacting on the sea surface temperature. Preliminary results obtained in this paper show that a slight but significant cooling in the temperature of the sea upper layer (of the order of $1\,^\circ$C every 50 years) could be happening since some decades ago at the northern Patagonian continental shelf waters.


\end{abstract}

\section{Introduction}
\label{intro}

Changes in wind due to global warming may have large geophysical impacts (McInnes et al. 2011). A number of changes in atmospheric processes have been reported at the Southwestern Atlantic Ocean. For instance, Barros et al. (2000) found that the western border of the South Atlantic High and the atmospheric circulation over Southeastern South America have slowly shifted towards the south during the last decades. Changes in wind speed have a significant impact on storm surge and wind wave climates at the Southeastern South America continental shelf (see, for example, D'Onofrio et al. 2008; Dragani et al. 2010; Codignotto et al. 2012; Dragani et al. 2013). In addition, wind speed changes play a fundamental role in the spatial patterns of sea surface temperature warming, the global hydrological cycle (through evaporation) and  in the regional distribution of sea level rise.

An increasing in the wind speed has also an important impact on the mean depth of the mixed layer of the shelf seas (Huang et al. 2006). The vertical structure of the water column is the result of ongoing competition between the buoyancy inputs due to surface heating and freshwater input, on the one hand, and stirring by the tides and wind stress, on the other. Variations in mixed-layer depth affect the rate of exchange between the atmosphere and deeper ocean, the capacity of the ocean to store heat and carbon and the variability of light and nutrients to support the growth of phytoplankton. However, the response of the Southern Ocean mixed layer to changes in the atmosphere is not well known (Sall\'ee et al. 2010).  On the Patagonian continental shelf the dominant buoyancy is mainly due to the seasonally varying surface heat ({Gue\-rre\-ro} and Piola 1997) because the rainfall is very scarce. During the winter months, when heat is lost from the surface, the buoyancy term contributes to stirring by increasing surface density and making all or part of the water column convectively unstable. As a result the shelf waters are vertically well mixed during the winter months. This vertically mixed regimes continues until the onset of positive heating at, or close to, the vernal equinox (around  September 21st in the southern hemisphere) after which the increasing input of positive buoyancy tends to stabilize the water column. Whether or not the water column stratifies depends on the relative strengths of the surface heating and the stirring due to frictional stresses imposed at the bottom boundary by the tidal flow and at the surface by wind stress. Huang et al. (2006) studied the decadal variability of wind-energy input to the world ocean and found that this energy varied greatly on inter-annual to decadal time scales. In particular, they showed that it has increased $12\%$ over the past $30$ years, and that the inter-annual variability mainly occurs in the latitude bands between $40^\circ-\,60^\circ$S. If that is the case, the deepening of the mixed layer would have produced a larger mixing of surface water and, consequently, a slight reduction of the sea surface temperature in this region of the South Western Atlantic Ocean.

\begin{figure}[!h]
\begin{center}
\includegraphics[width=\textwidth]{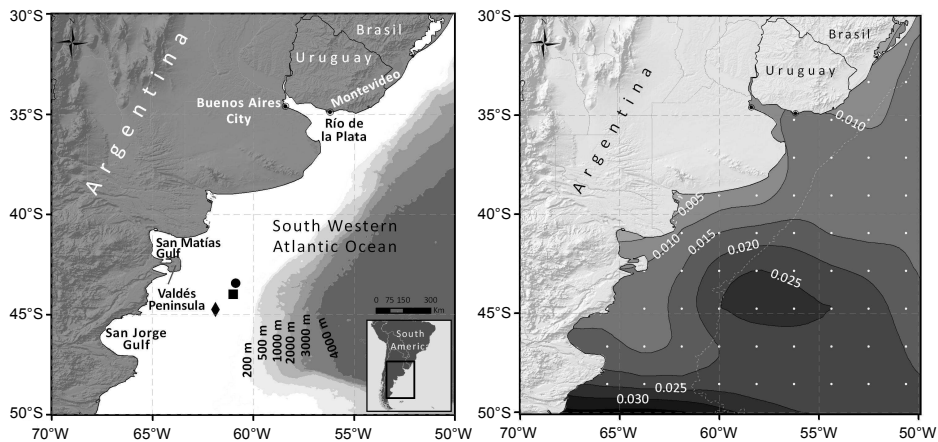}
\caption{{\it Left:} Relevant points in the analysis: NR1 node (diamond), TPXO tidal point (square), and site where the water temperature profiles were measured (circle). {\it Right:} Annual trend of sea surface wind speed (m$\, $s$^{-1}$). White squares show NR1 grid nodes; 200 m contour is also indicated in white dashed line}
\label{figElSitio}
\end{center}
\end{figure}

Direct observations collected over the Patagonian continental shelf waters (Argentina) indicate that during the 90's, winds were $20\%$ stronger than during the 80's and that the northwest direction was more frequent (Gregg and Conkright 2002). In the region of study in this work (Fig. \ref{figElSitio} left panel), a maximum wind speed trend  greater than $2.5$ cm$\, $s$^{-1}$/yr was estimated from NCEP/NCAR I reanalysis (NR1) at the border of the Patagonian continental shelf (at $44^\circ$S $60^\circ$W, approximately; Fig. \ref{figElSitio} right panel).

The aim of this paper is to investigate a possible deepening of the mixed layer at a selected point of the Patagonian shelf where atmospheric and oceanographic data are available (Fig. \ref{figElSitio} left panel) approximately at $44^\circ$S $61^\circ$W.  This study was carried out by means of a 1-dimensional numerical model (Sharples et al. 2006; Sharples 1999) developed to study the physical structure of the upper layer of the ocean at coastal and shelf seas. This model was forced using atmospheric data from NR1 reanalysis and tidal constituents from TPXO 7.2 global model. Numerical results were conveniently validated using temperature profiles observed during Summer and Winter at the Patagonian continental shelf. This region is ecologically and economically important. One of the most important economic activities in the area is fisheries, while its coastal marine fauna is practically unique in the world. The Patagonian shelf is characterized by a smooth slope and scarce relief features (Parker et al. 1997). The shelf broadens from north to south, ranging from 170 km at $38^\circ$S to more than 600 km south of $50^\circ$S. The main source of the shelf water masses is the subantarctic water flowing from the northern Drake Passage, through the Cape Horn Current (Hart 1946) between the Atlantic coast and the Malvinas/Falklands Islands, and the Malvinas/Falklands Current in the eastern border of the shelf (Bianchi et al. 2005). The fresh water sources of the shelf are the small continental discharge and the low salinity water outflowing from the Magellan Strait. South of $41^\circ$S, the shelf width is close to one quarter of the semidiurnal tide wavelength, leading to favorable conditions for resonance (Piola and Rivas 1997). The tidal amplitude in the Patagonian shelf is one of the highest in the world ocean (Kantha et al. 1995), and tidal currents are very energetic (Simionato et al. 2004).

\section{Data}
\label{sec:data}

\begin{figure}[!h]
\begin{center}
\includegraphics[width= \textwidth]{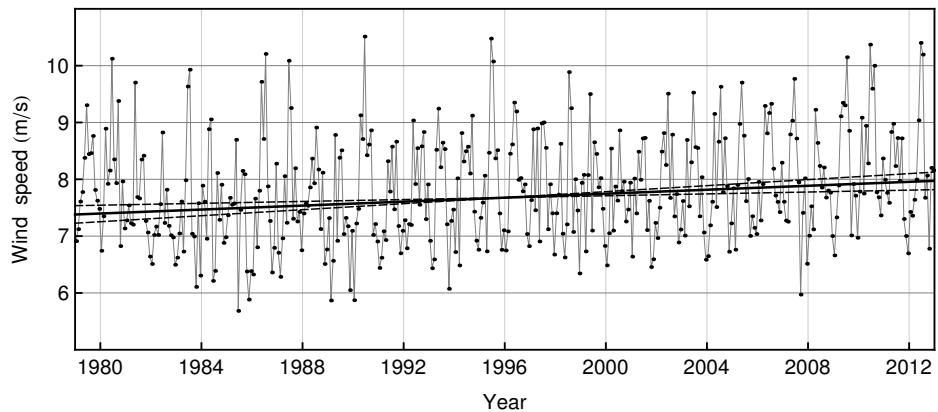}
\caption{Wind speed trend in the 1979-2012 studied period of time. Each point represents the average wind speed of one month with data taken every 6 hours.  The Kendall-Theil algorithm yields a wind speed variation of  $1.7 \pm 0.9$ cm$\,$s$^{-1}$/yr at a 95\% confidence level.  (The same result is obtained using a regular linear fit.) The solid line represents the average wind speed increase and the dashed lines its 95\% confidence interval}
\label{wind}
\end{center}
\end{figure}

Surface (10 m height) zonal and latitudinal wind components, surface (2 m height) air temperature, surface atmospheric pressure,  surface (2 m height) specific humidity, and cloud coverage from NR1 (period: 1979-2012) at a grid node located at $44.7611^\circ$S $61.8750^\circ$W were used as atmospheric forcing of the model. The result of NR1 is a set of gridded data (Global T62 Gaussian grid) with a temporal resolution of 6 h. Data before 1979 were not included in this study due to known deficiencies of the reanalysis prior to satellite era, particularly in data-sparse regions such as the high-latitude Southern Hemisphere (Jones et al. 2009; Bromwich and Fogt 2004). The main advantages of this reanalysis are its physical consistency and high temporal coverage. Full details of the NR1 project and the dataset are given in Kalnay et al. (1996) and discussions on its quality in the Southern Hemisphere can be found in Simmonds and Keay (2000), among others. The raw analysis of the NR1 data shows no significant trend, except in the wind.  In Fig. \ref{wind} we plot the wind speed evolution at the NR1 grid node for the studied period of time and find a significant increment of $1.7 \pm 0.9$ cm$\, $s$^{-1}$/yr at a 95\% confidence level.

Current tidal constituents for five primary harmonic constituents (M2, S2, N2, K2, K1) at the grid node located at $44^\circ$S $61^\circ$W were obtained from TPXO 7.2 model (http://volkov.oce.orst.edu/tides/global.html) to get a realistic representation of the tidal dynamics. TPXO 7.2 is a recent version of a global model of ocean tides, which best-fits the Laplace Tidal Equations and along-track averaged data from TOPEX/Poseidon and Jason (on TOPEX/POSEIDON tracks since 2002) obtained with OTIS. The methods used to compute the model are described in detail by Egbert et al. (1994) and further by Egbert and Erofeeva (2002).

We used two temperature quasi-continuous vertical profiles to validate the numerical simulations. Both profiles were collected within the framework of the Coastal Contamination, Prevention and Marine Management project, which was part of the scientific agenda of the United Nations Development Program. Cruises were carried out on board the "Puerto Deseado" oceanographic vessel, in March and September 2006. In each cruise, hydrographic stations were carried out along cross-shelf sections spanning the shelf from near-shore to the western boundary currents, between $38^\circ$ and $55^\circ$S (Charo and Piola 2013).  For the purpose of this work, we used data collected at $43.4438^\circ$S $60.8673^\circ$W (March 29th, 2006), and $43.4473^\circ$S $60.8663^\circ$W (September 9th, 2006). Water depth at these locations is $L=97$ m.

\section{Numerical study of water column temperature profile and thermocline definition}
\label{sec:numerical}
In this section we describe the methodology that, from the input meteorological variables, yields the thermocline and mixed layer features.  We validate the model using available {\it in situ} data.

\subsection{{\tt S2P3} Numerical model}
\label{s2p3}

The program S2P3 is a 1-dimensional (vertical) coupled physical-biological numerical model. It uses meteorological data and tidal currents to simulate the 365-day evolution of several physical and biological variables within the water column. A detailed description can be found elsewhere (Sharples et al. 2006; Sharples 1999).

By default, each run of the model starts on January 1st (day 1), when, for the northern hemisphere,  the water column is well mixed and has a uniform initial temperature. This initial water temperature must be provided by the user. Some minor changes were implemented to adapt the model to the southern hemisphere:
\begin{itemize}
\item First, the calendar must be shifted six months, which means that days must be numbered from 1 to 365 starting on July 1st. Our convention will be to name each of these 365-day periods according to the year in which they begin.  For instance, 1995 must be understood as the period of 365 days beginning on July 1st, 1995 and ending on June 30th, 1996 (leap years will end on June 29th). We will call this {\it the winter calendar}. Our simulations span a period of 33 years, beginning on July 1st, 1979 and ending on June 29th, 2012.
\item Second, in order to reproduce the correct solar irradiance the sign of the latitude of the study point should be changed. Accordingly, in our simulations the input latitude was $44^\circ$N.
\item Third, to preserve the correct relative directions between the Coriolis force, wind stress, and tidal current the latitudinal components of the tidal current and wind velocity should be inverted.
\end{itemize}
The water depth at the site of the simulation is 97 m. The numerical model divides the water column in $n$ cells of equal thickness. Trade-off between resolution and computation time must be considered. After some preliminary tests, we carried out the bulk of our simulations using 100 cells, for which smooth and consistent profiles resulted.

The available meteorological data from NR1 had a resolution of 6 hours, while the input for the numerical model requires only one daily value for each variable. For the surface air temperature, atmospheric pressure and specific humidity, we used the average of the four daily values from NR1. For cloud coverage, we used the average of the two daytime values. Cloud coverage is also important during nighttime, because it affects heat losses. In any case, the difference between daytime and nighttime  cloud coverage proved to be less than $5\%$ in average. As for the wind speed and direction, for each day, starting with the four daily values available from the NR1 data, we computed a weighted average, univocally defined by the following condition: that the wind stress computed from this weighted average equals the average of the stresses computed from the four daily values of the wind speed and direction.
As the initial water temperature for the model we used the measured winter temperature of the sea bottom, 6.8 $^\circ$C.  All other parameters were left in the {\tt S2P3} original tune.

The model outputs hourly vertical profiles for several variables, including water temperature (instantaneous value) and turbulent kinetic energy per mass unit (averaged over one-hour periods). We used these two variables to define the thermocline boundaries and the mixed layer depth.

\subsection{Model validation}
\label{validation}

In order to validate the model, we used available temperature profiles for two reference dates: September 9th and March 29th, 2006. These dates correspond to days 70 and 271, respectively, in the winter calendar explained in Sect.~\ref{s2p3}.  Therefore, we expect that at day 70 the temperature profile is relatively more sensitive to meteorological fluctuations and, on the other hand, at day 271 the summer is ending and the temperature profile has reached a good contrast and shaped a better defined mixed layer, thermocline, and deep water regions which are more robust to fluctuations.

\begin{figure}[!h]
\begin{center}
\includegraphics[width=0.75\textwidth]{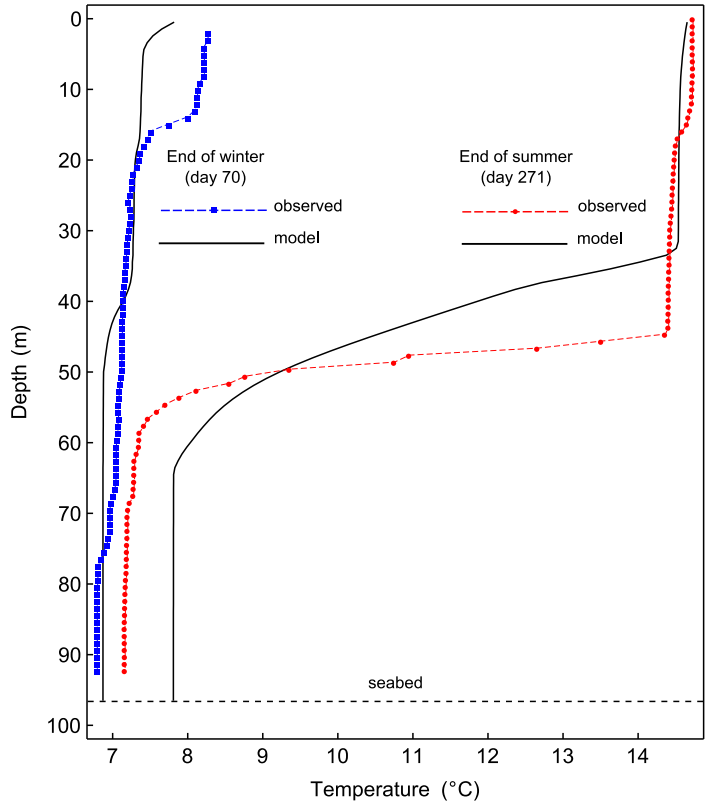}
\caption{Validation plots for the numerical model.  The solid black lines represent the outcomes of the numerical model, whereas dashed lines represent {\it in situ} measurement of the water temperature profile for two reference dates}
\label{validacion}
\end{center}
\end{figure}

In Fig.~\ref{validacion} we have plotted the measured temperature profile and the one predicted by the {\tt S2P3} software using the NR1 meteorological data for the mentioned reference dates.  As it can be seen in the figure, the agreement is very good for both reference dates.

\subsection{Study of temperature profiles: mixed layer and thermocline definition}

Having validated our numerical model for the temperature profile, we are interested at this point in defining the mixed layer for the water column, in order to study its properties and evolution at the analyzed location from 1979 to 2011.

\begin{figure}[!h]
\begin{center}
\includegraphics[width= \textwidth]{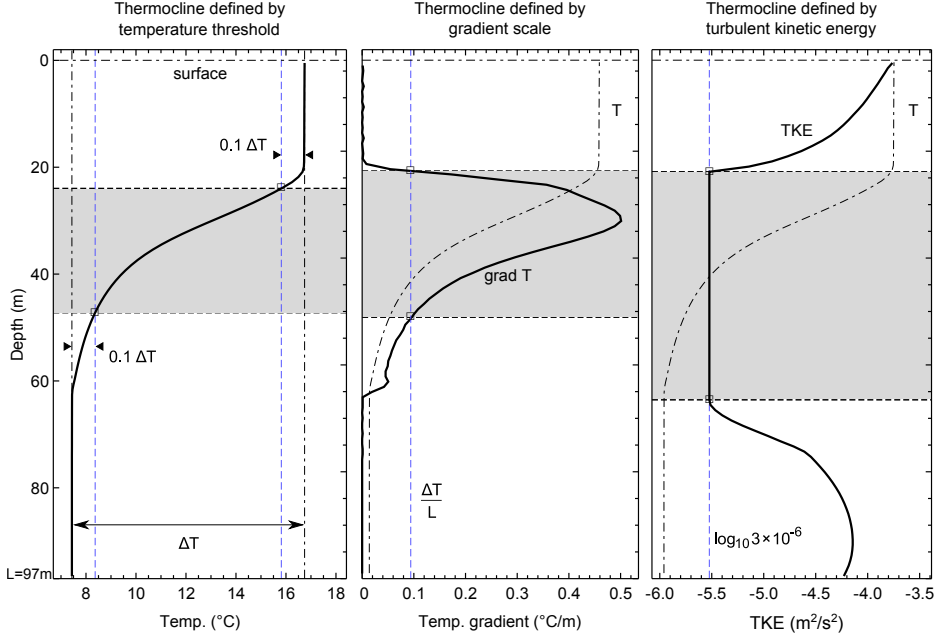}
\caption{Outcome of the three thermocline definitions (see text) for the same summer day at noon.  The white upper region corresponds to the mixed layer, the shaded region to the thermocline, and the bottom region to the deep water region.  All three definitions, although quite different, yield a similar mixed layer region.  $\Delta T$ is the temperature difference between surface and seabed, and $L = 97$ m is the depth of the seabed}
\label{3def}
\end{center}
\end{figure}

There are many algorithms which yield differently defined mixed layer  and thermocline boundaries.  These different definitions yield similar mixed layer or thermocline upper boundary for good contrast temperature profiles.  The lower thermocline boundary, on the other hand, may yield some differences due to its relative smoothness in contrast to the upper boundary.  In this work, we use the following three algorithms to define the mixed layer and thermocline boundaries:
\begin{itemize}
\item {\it Temperature threshold: } We define the thermocline upper boundary as the depth at which the temperature decreases $0.1\Delta T$ with respect to the surface, where $\Delta T$ is the temperature difference  between the surface and the seabed.  Analogously, the lower boundary is defined as the depth at which the temperature increases by $0.1\Delta T$ with respect to the seabed. (See left panel in Fig.~\ref{3def}.)
\item {\it Gradient scale:} The upper and lower boundaries are defined as the two depths at which the temperature gradient equals the mean temperature gradient between the surface and the seabed. (See central panel in Fig.~\ref{3def}.)
\item {\it Turbulent kinetic energy (TKE):} We define the thermocline as the region where TKE (per unit of mass) is less than or equal to a critical value. In fact, in the numerical model there is a minimum, fixed value for TKE, equal to $3\times 10^{-6}$ m$^2$ s$^{-2}$. The thermocline is the region where TKE takes this minimum value. With this definition, the mixed layer is the region next to the surface in which active turbulence is present. (See right panel in Fig.~\ref{3def}.)
\end{itemize}
In Fig.~\ref{3def} we show the three different definitions for a typical summer day at noon.  As it can be seen in the figure, the agreement for the mixed layer is very good, whereas the lower thermocline boundary may have some discrepancies.  In any case, since in this work we focus on the mixed layer features we will proceed using simultaneously all three thermocline definitions and verify if our conclusions are independent of which one we use.  We also note that there is relatively little hour-to-hour variation within a given day. In the rest of this paper the mixed layer depth for each day is given by averaging the 24 hourly depth values computed from the hourly outputs of the numerical model.

\begin{figure}[!h]
\begin{center}
\includegraphics[width= \textwidth]{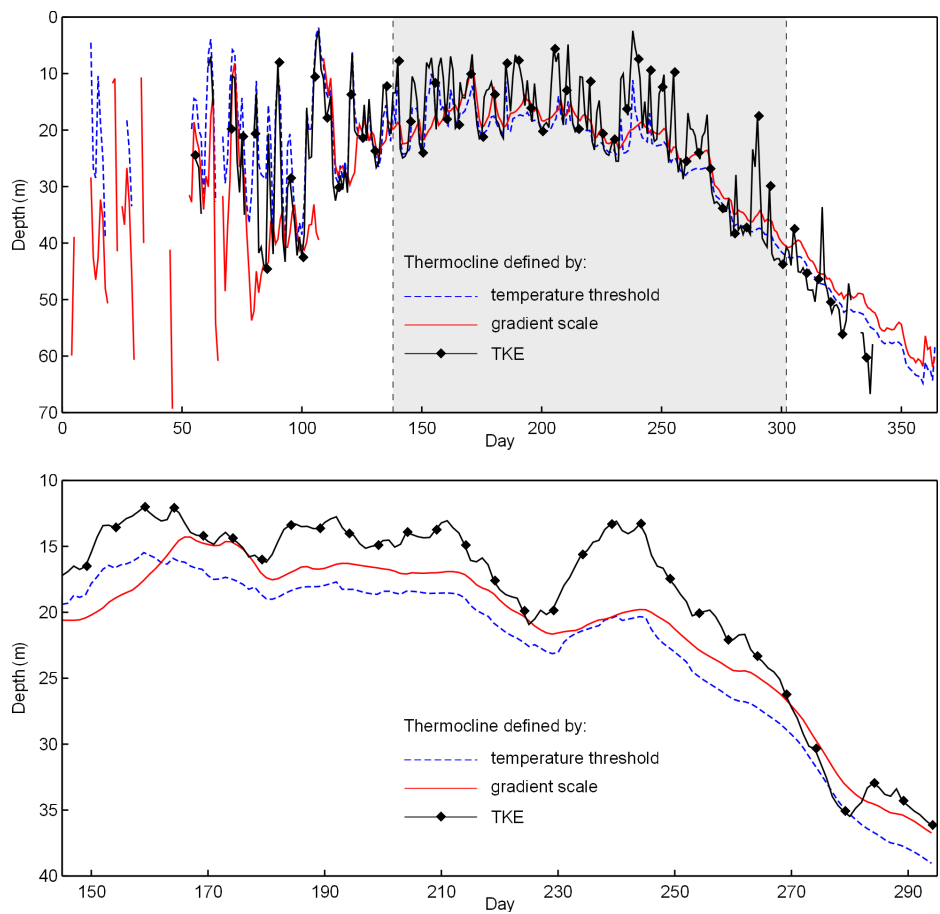}
\caption{{\it Top:} Mixed layer depth for a complete year using the three thermocline definitions.  {\it Bottom:} selected time window between days 138 and 302, when the contrast reached in water temperature yields an outcome more stable to weather fluctuations.  In the bottom plot a 15-day moving average has been applied to soften the curves}
\label{thermoclineyear}
\end{center}
\end{figure}

To study the stability along the year of the three mixed layer depth definitions, we show in Fig.~\ref{thermoclineyear} (top panel) the direct outcome of each definition for each day of a typical year.  In the bottom panel of the figure, we select the relatively stable period between days $138$ and $302$ and do a moving average of 15 days for the mixed layer depth.  We see that in this selected period there is an agreement of the  three definitions and, since there is a good contrast in the temperature profile, the mixed layer seems stable upon weather fluctuations.

In the following sections we study the mixed layer depth and its time evolution along the years for the selected time window between days $138$ and $302$ of the winter calendar.

\section{Analysis of mixed layer depth and its trend}
Using the previous mixed layer definitions, as well as the time period in the year in which its depth is relatively robust to meteorological fluctuations, we study in this section the mixed layer depth evolution in the 1979-2011 33-year period.  For the sake of concreteness we show the results in this section using the TKE mixed layer definition.  In any case, we stress that similar results are obtained when the other two mixed layer definitions are used. We analyze the mixed layer depth evolution for six reference days within the 138-302 days period. To diminish the impact of weather fluctuations, for each of these reference days we average the mixed layer depth in a interval of the surrounding same 15 days in each year --i.e., given a day we include in the average the previous and following 7 days. This yields for each reference day a set of 33 points --one for each year-- with an average mixed layer depth.  Since each year has a fluctuating different weather, we do not expect to see a pattern in this set of 33 points.  However, if these fluctuations lie on a smoothly changing function, we can expect that a linear fit in each of these sets gets rids of the fluctuations and gives us a hint on the first derivative of this smoothly changing function for each of the reference days.

\begin{figure}[!h]
\begin{center}
\includegraphics[width= \textwidth]{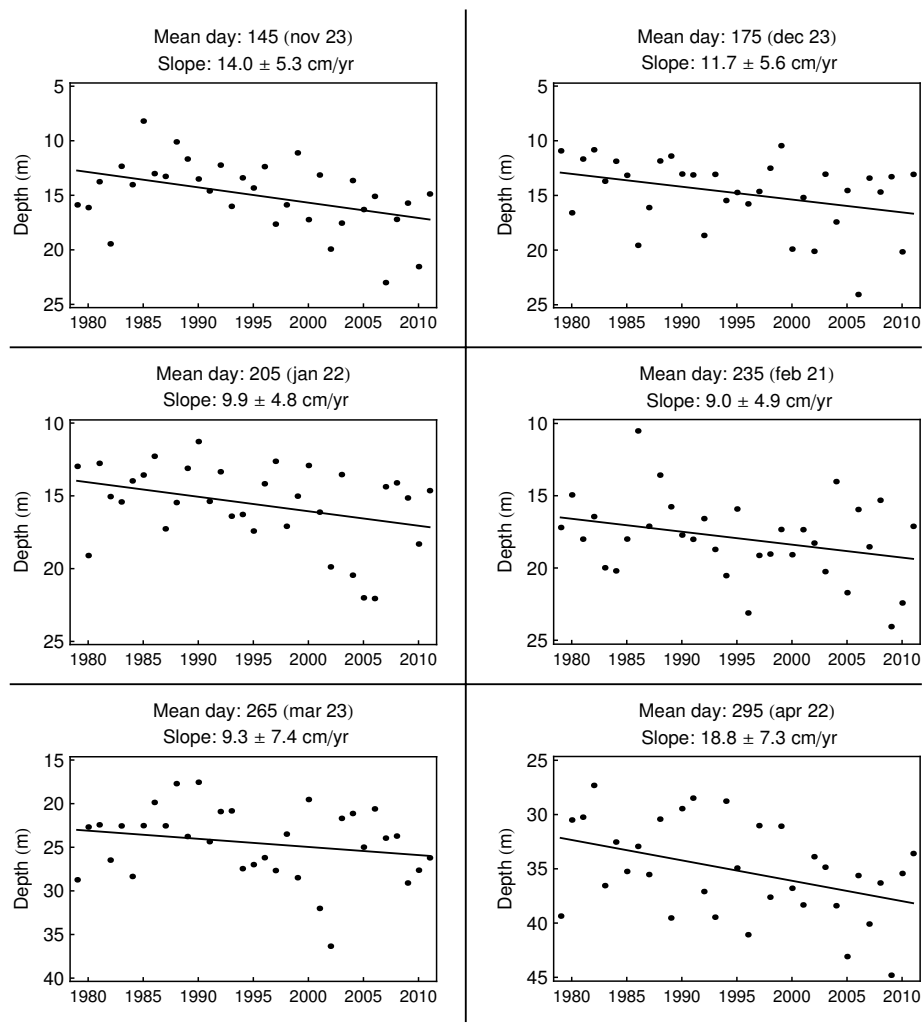}
\caption{Fit to the mixed layer depth as a function of the years for six reference days within the time window determined in Fig.~\ref{thermoclineyear}.  Each point represents the average of the mixed layer depth in the 15 days surrounding the given day}
\label{sixplots}
\end{center}
\end{figure}

In Fig.~\ref{sixplots} we show the results of this analysis and we found that for all six reference days the linear fit yields a positive (in depth) slope with a significance that ranges from approximately one to three standard deviations of the fit.

\begin{figure}[!h]
\begin{center}
\includegraphics[width=\textwidth]{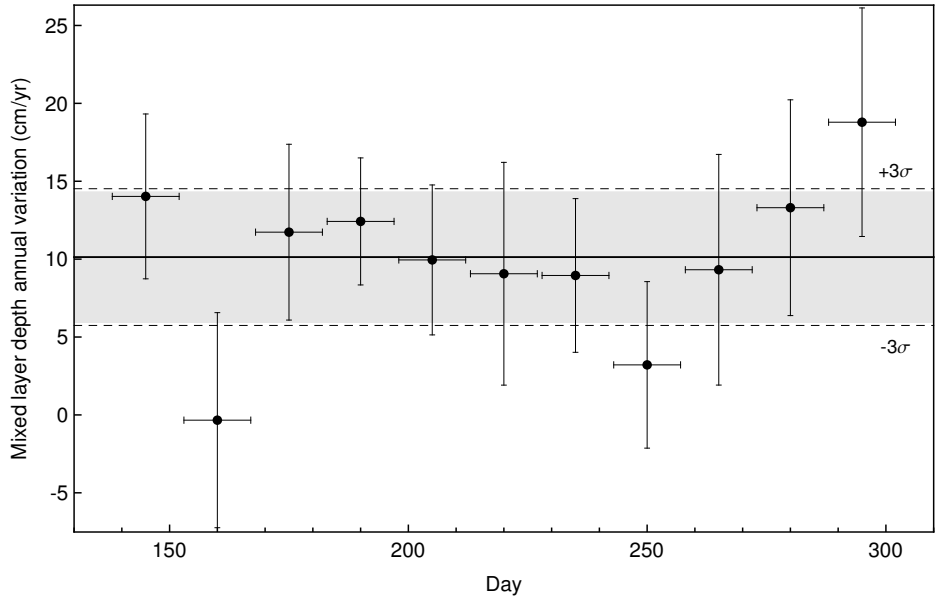}
\caption{Linear fit to the mixed layer depth annual variation for 11 equally long bins in the 138-302 days time window during the 1979-2011 period.  The outcome of the fit is $10.1\pm 1.4$ cm/yr which is more than 7 standard deviations different from zero}
\label{fixteen}
\end{center}
\end{figure}

Motivated by this first set of results, which shows a clear positive trend in the mixed layer depth, we proceed to find the average trend of this depth for the full time window in which the mixed layer depth is relatively stable to weather fluctuations.  In this sense, we take 11 bins of 15 days each, from day 138 to day 302, and compute the slope of the mixed layer depth as described in the previous paragraph.  To have the full average variation, we do an average on these 11 slopes, and obtain a final result of $10.1\pm 1.4$ cm/yr (see Fig.~\ref{fixteen}).  This result, being one of the most important in the article, would indicate a clear trend of enlargement of the mixed layer along the years by more than $\sim 7$ standard deviations.  Incidentally, it is worth to notice at this point that the other two mixed layer definitions yield similar results: $11.1 \pm 1.2$ cm/yr (temperature threshold) and $10.6 \pm 1.1$ cm/yr (gradient scale).

\section{Discussion}

\begin{figure}[!h]
\begin{center}
\includegraphics[width= \textwidth]{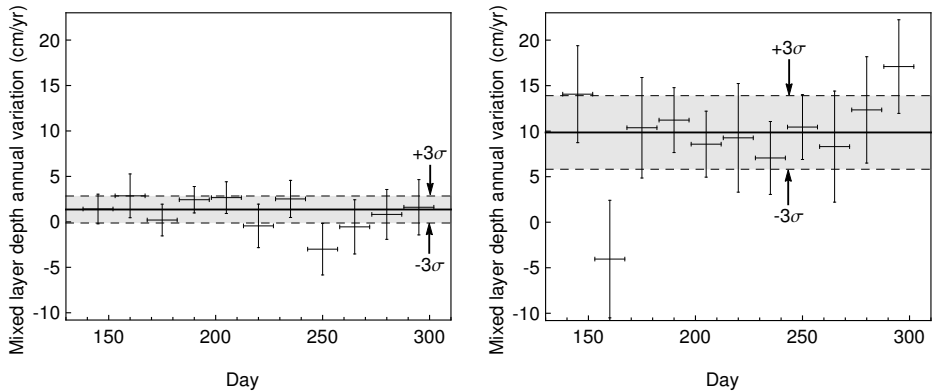}
\caption{Same analysis as in Fig.~\ref{fixteen}, but for two thought scenarios.  {\it Left:} in all years the wind has been set to its 1979 value leaving all other variables in its real values.  {\it Right:} the wind in each year has been left in its real value, but all other variables have been set to their 1979 values for all years. These plots suggest that the wind would be the major cause of the increase trend found in Fig.~\ref{fixteen}}
\label{1979}
\end{center}
\end{figure}

\begin{figure}[!h]
\begin{center}
\includegraphics[width= \textwidth]{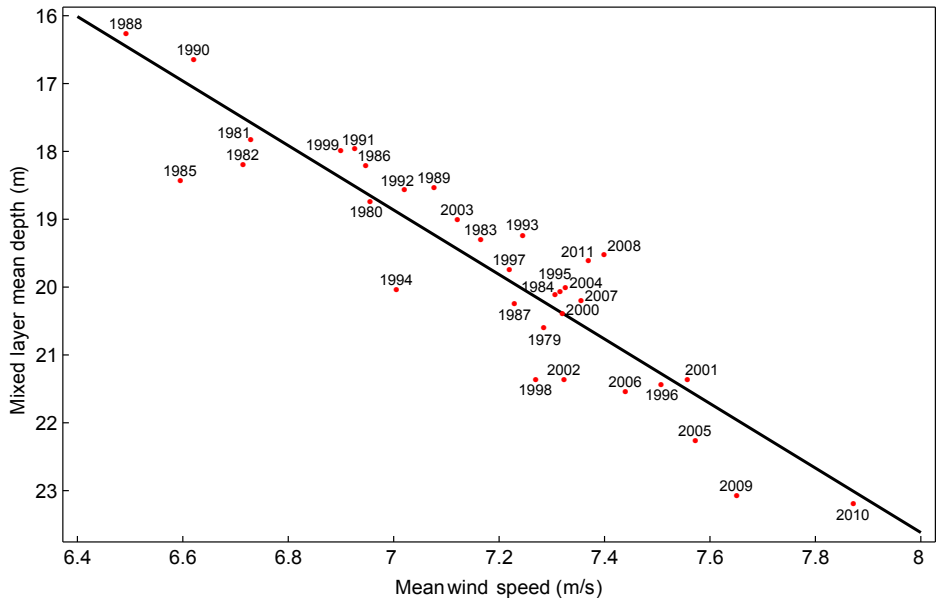}
\caption{Scattered plot for the mean wind speed versus the mixed layer mean depth during the stratification season for each of the studied years.  The correlation between these two variables can be seen by a linear fit on them which yields the solid line in the plot with a slope equal to $4.8 \pm 0.4$ cm/(cm$\,$s$^{-1})$.   Pearson correlation coefficient is $0.916$.  Notice the accumulation of recent years in the bottom right of the figure}
\label{scattered}
\end{center}
\end{figure}

Given the results obtained in the previous section, we question whether there is a main feature in the meteorological data that may be generating the enlargement of the mixed layer.  In order to explore this question, we have performed the following numerical experiments.  In a first thought scenario, we have used the 1979 wind meteorological data for all the 1979-2011 period, leaving the rest of the meteorological variables in its original values.  On the other hand, in a second though scenario, we have used the original wind data for each year, but the rest of meteorological variables have been left with the 1979 data for all the years in the 1979-2011 period.  We have then performed the same analysis that yields to Fig.~\ref{fixteen}, but for these two thought scenarios, and reached the results shown in Fig.~\ref{1979}.  We have found that in the first scenario there is no significant trend of the mixed layer depth ($1.4 \pm 0.5$ cm/yr), whereas in the second scenario we retrieve a very similar result to the one with the actual meteorological conditions ($9.9 \pm 1.3$ cm/yr). These results would be indicating that the change in the wind is the main cause of the increase trend in the mixed layer depth.  To strength this conclusion, we have plotted in Fig.~\ref{scattered} the mean wind speed versus the mixed layer mean depth during the stratification season and we confirm not only a correlation between these two variables, but also a trend of recent years to have deeper mixed layer and stronger winds.


In order to have a better understanding on the influence of the meteorological data on the model for the water column temperature, it is useful to notice some relevant features of the data, the model and the obtained results.  As mentioned in Sect.~\ref{sec:data} there is no significant trend in the temperature data, moreover, in the model the incoming flux of solar radiation is fixed. Therefore, neglecting wind energy that may be converted into heat, the system does not have an inter-annual change in energy deposited in the ocean.  As a matter of fact, the day of the year for which the water column has accumulated the maximum amount of energy (computed assuming constant density and specific heat), averaged on all the years, is day $260 \pm 2$, with no annual trend in this result.  These observations suggest that the real effect of the wind in enlarging the mixed layer is a greater mixture of the water column.  To support this hypothesis, we observe that the results of the numerical model show that the maximum annual sea surface temperature would be having a slight decrease at a rate of $1.2 \pm 0.6 \,^\circ$C every 50 years. This trend is in very good agreement with linear trend estimated by Richard et al. (2012) from UK Meteorological Office Hadley Centre's SST dataset version 2 (HadSST2, $5^\circ\! \times 5^\circ$ regular global grid, period 1958-2002). This dataset is only based on observations and does not interpolate surface temperature over the regions that could not be documented. It is thus thought to minimize statistical artifacts in the southern high-latitudes, where {\it in situ} measurements are rare. Richard et al. (2012) obtained a negative annual trend of approximately $-0.2\,^\circ$C/decade (figure 6 of their paper) and monthly trends lower than $-0.2\,^\circ$C/decade, from February to May (figure 7 of their paper), at the northern Patagonian continental shelf.

\section{Conclusions}

A possible deepening of the ocean mixed layer at the northern Patagonian continental shelf (Fig. \ref{figElSitio} left panel) was investigated by means of a 1-dimensional (vertical) numerical model (Sharples et al. 2006; Sharples 1999) forced by atmospheric data from NCEP/NCAR I (NR1) and tidal current constituents from TPXO 7.2 global model. The performance of the model was analyzed by means of the comparison between simulated and observed temperature profiles and it was concluded that the model is fairly reliable to simulate the vertical thermal structure at the region (Fig. \ref{validacion}). Model validation and numerical simulations and experiments were carried out at a location of the Patagonian continental shelf where significant positive wind speed trends were previously detected (Fig. \ref{figElSitio} right panel and Fig. \ref{wind}). Three different criteria were tested to determine objectively the thickness and evolution of the simulated mixed layer, showing no appreciable variation (Figs. \ref{3def} and \ref{thermoclineyear}).

From a long term numerical simulation (1979-2011) it was found that the mixed layer thickness showed a significant and positive trend which seems to be noticeable variable along the year (Fig. \ref{sixplots}). In addition, the mean mixed layer depth trend computed from the 33 simulated years, which could be considered as a representative value of the mixed layer thickness trend for the region, resulted equal to $10.1 \pm 1.4$ cm/yr (Fig. \ref{fixteen}). In order to analyze the sensitivity of these results some numerical experiments were carried out. A numerical experiment was carried out using the prescribed atmospheric input (NR1 data from 1979 to 2011), except the wind speed which was set to values corresponding to 1979. As a result, no significant trend was obtained in this case (Fig. \ref{1979} left panel). On the contrary, another numerical experiment was carried out forcing with the prescribed wind speed (NR1 data from 1979 to 2011) but setting the additional atmospheric input to values corresponding to 1979. Accordingly, almost the same previous significant positive trend  was obtained in this experiment (Fig. \ref{1979} right panel). This clearly supports that the reported wind speed increasing could be considered as the main responsible of deepening of the ocean mixed layer at the Patagonian continental shelf. This is also shown in the Fig. \ref{scattered}, where an evident correlation between a deeper mixed layer and a stronger wind can be clearly appreciated.

Finally, our results suggest that a possible enlarging of the mixed layer would produce a deeper mixture of the upper water column and, in consequence, the maximum annual sea surface temperature could be decreasing at a rate of about $1\,^\circ$C every 50 years. This negative trend is in very good agreement with linear trend estimated by Richard et al. (2012) from HadSST2 database (period 1958-2002) at the northern Patagonian continental shelf. Moreover, we provide qualitative arguments to understand this lowering in the ocean surface temperature.

\begin{acknowledgements}
This work is a contribution to the projects PIP (2012-2014) 00176 from CONICET and PICT-2011-0359 from the ANPCyT.
\end{acknowledgements}

\end{document}